\documentclass[twocolumn, pre, showpacs, english, preprintnumbers, amsmath, amssymb, superscriptaddress, aps,longbibliography]{revtex4-2}
\usepackage[utf8]{inputenc}
\usepackage{graphicx}
\usepackage{xcolor}
\usepackage{appendix}

\let\oldepsilon\epsilon \let\epsilon\varepsilon \let\varepsilon\oldepsilon

\makeatother

\begin{document}

\title{
Anisotropic differential conductance of a mixed parity superconductor/ferromagnet  structure}

\author{Tim Kokkeler}
\email{tim.kokkeler@dipc.org}
\affiliation{Donostia International Physics Center (DIPC), 20018 Donostia--San Sebasti\'an, Spain}
\affiliation{University of Twente, 7522 NB Enschede, The Netherlands}

\author{Alberto Hijano}
\email{alberto.hijano@ehu.eus}
\affiliation{Centro de F\'isica de Materiales (CFM-MPC) Centro Mixto CSIC-UPV/EHU, E-20018 Donostia-San Sebasti\'an,  Spain}
\affiliation{Department of Condensed Matter Physics, University of the Basque Country UPV/EHU, 48080 Bilbao, Spain}

\author{F. Sebasti\'{a}n Bergeret}
\email{fs.bergeret@csic.es}
\affiliation{Centro de F\'isica de Materiales (CFM-MPC) Centro Mixto CSIC-UPV/EHU, E-20018 Donostia-San Sebasti\'an,  Spain}
\affiliation{Donostia International Physics Center (DIPC), 20018 Donostia--San Sebasti\'an, Spain}

\begin{abstract}
We study the electronic transport properties of a  superconductor (S)  with a mixed s+p-wave pairing attached to a ferromagnetic metal (F) and a  normal electrode (N) in an SFN configuration. Using the quasiclassical Green's function method, we compute the differential conductance $\sigma$ of the junction and demonstrate its dependence on the direction of the exchange field relative to the direction of the d-vector of the pair potential. If the p-wave triplet dominates the pairing, 
 the zero bias conductance depends on the relative direction between the triplet d-vector and the exchange field. In contrast, if the s-wave singlet dominates the pairing,  the zero bias conductance is isotropic with respect to the field direction. Furthermore, at zero temperature,  the zero bias conductance height can only take two values as a function of $r$, the parameter quantifying the relative amount of s- and p-wave pairing, with an abrupt change at $r=1$ when the superconductor goes from a singlet to triplet dominated ground state.
Moreover, we show that the relative amount of s- and p-wave pairing, can be estimated from the dependence of the finite bias conductance on the exchange field direction. Our results provide a way to characterize parity-mixed superconductors performing electrical measurements.
\end{abstract}

\maketitle

\section{Introduction}
Among the various types of unconventional superconductors, much attention has been paid to the study of  superconductors with triplet correlations \cite{mackenzie2017even,kallin2016chiral,linder2019odd,balian1963superconductivity,sigrist1991phenomenological,fu2008superconducting,Bergeret-2005,chiu2021observation}. These correlations can be  induced either via the proximity effect by 
combining superconductors with other materials \cite{fu2008superconducting,Bergeret-2005}, or they may exist in bulk superconductivity, for example in uranium based ferromagnetic superconductors \cite{aoki2019review, saxena2000superconductivity,aoki2001coexistence,hardy2005p,huy2007super,ran2019nearly}.

Most works focus on  superconductors which have inversion symmetry, that is, in which the parity of the pair potential is either even or odd. However, in the past few decades superconductors have been discovered whose underlying crystal structure  lacks inversion symmetry \cite{bauer2012non,bauer2004heavy,amano2004superconductivity,akazawa2004pressure,togano2004Superconductivity,tateiwa2005novel,kimura2005pressure,sugitani2006pressure,honda2010pressure,SETTAI2007844,Bauer2010Unconventional, xie2020captas,yang2021spin}. In such superconductors  parity-mixed superconductivity may arise \cite{bauer2012non}. 
Non-centrosymmetric superconductors have interesting applications, for example,  they are very suitable for superconducting diodes due to the inversion symmetry breaking \cite{wakatsuki2017nonreciprocal, narita2022field}.

An important issue is the determination of the pair potential. There have been many efforts to explore restrictions on the possible pair potentials and to predict properties of inversion-symmetry broken superconductors \cite{levitov1985magnetostatics,edel1989characteristics,gor2001superconducting,mineev2004superconductivity,frigeri2004superconductivity,frigeri2006phenomenological,yanase2008superconductivity,gentile2011odd,bauer2012non, Mineev_2017,borkje2006tunneling,mineev2011magnetoelectric}. Still though, in general it is difficult to determine the type of unconventional pairing.  
Examples of efforts include using NMR \cite{samokhin2005nmr,hayashi2006nuclear,aso2007incommensurate, Pustogow-2019,Aoki-2019} or measuring the critical field for different directions of an applied magnetic field \cite{Hardy-2005,samokhin2008effects,samokhin2008upper}, to identify spin-triplet pairing. s+p-wave pairing is predicted to be, under certain conditions,  the most stable pairing, for example  in  $\text{CePt}_{3}\text{Si}$  \cite{yanase2008superconductivity}.
 There are also theoretical suggestions to explore the proximity effect of unconventional superconductors on normal materials \cite{iniotakis2007andreev,eschrig2010theoretical,annunziata2012proximity,rahnavard2014magnetic, mishra2021effects, kokkeler2022phys}. 
However, for many materials, the results are not conclusive.

In this work, we explore non-equilibrium electronic transport through a superconductor/ferromagnet/normal metal (SFN) junction,  to reveal properties of the parity-mixed pair potential.  We focus on the simplest type of a parity-mixed pair potential, the s+p-wave superconductor,  with  a helical p-wave pairing. 
We calculate  the differential conductance $\sigma$ of the junction shown in Fig. \ref{fig:Setup} and investigate the dependence of $\sigma$ on both the amplitude and direction of the intrinsic exchange field of the F metal.
We first focus on the zero bias conductance.  It shows a peak when  $\Delta_{t}>\Delta_{s}$.  We find that the height of the zero bias conductance peak (ZBCP)   remains unchanged for  exchange fields that are perpendicular to the direction of transport. In contrast, when the exchange field is parallel to the d-vector, the differential conductance peak shifts to finite  voltages and the zero bias conductance is suppressed. 
Thus, for large exchange fields only a broad dome-like shape remains.
The zero bias conductance   varies monotonically as a function of the angle between d-vector and exchange field.

We also show  that the angular dependence of the differential conductance for nonzero voltages can be used to determine the mixing parameter, the relative strength of the singlet and triplet components of the pair potential. If $\Delta_{s}>\Delta_{t}$, a long junction with $E_{\text{Th}}<\Delta_{0}$,  can be used for this purpose. Here $E_{\text{Th}}=D/L^2$ is the Thouless energy,  $L$ and $D$ are  the length and diffusion coefficient of the   the F link respectively, and  $\Delta_0$ the amplitude of the gap. 
If $\Delta_{t}>\Delta_{s}$,  a short junction with $E_{\text{Th}}\sim\Delta_0$ is more suitable for the determination of the mixing parameter.
We also find that the exchange field dependence of the zero bias conductance in both the long and short junctions is independent of the exact ratio between $\Delta_{s}$ and $\Delta_{t}$, it is fully determined by whether the singlet component or the triplet component is dominant.
Thus, with the proposed setup the pair potential of an s+p-wave  superconductor can be fully characterized by  electrical measurements.

The work is organized as follows.  In  section \ref{sec:equations} we introduce the equations used to describe the system and the boundary conditions at the interfaces between different materials. In section \ref{sec:results}, we present our results for  the differential conductance.
 We also show how the differential conductance can be used to reveal the mixing parameter between the singlet and triplet amplitudes. Section \ref{sec:conclusion} is devoted to a discussion of the results and an outlook. Throughout the paper we work in units with $\hbar = k_{B} = 1$.

\section{The Model}\label{sec:equations}
\begin{figure}
    \centering
    \includegraphics[width = 8.6cm]{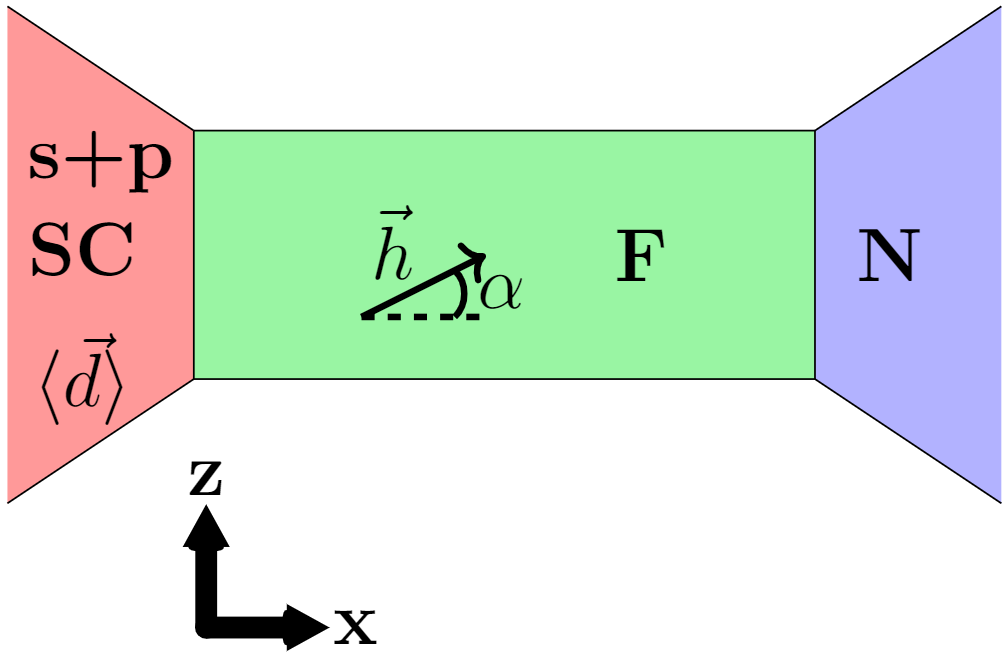}
    \caption{A schematic of the SFN junction.  The superconductor is a s+p  mixed-parity superconductor.   A voltage is applied to the normal metal electrode (N) to drive currents through the junction. The differential conductance is calculated as function of the direction of the exchange field $\vec{h}$ in the ferromagnetic bar (F).}
    \label{fig:Setup}
\end{figure}

We consider a  ferromagnetic metal of mesoscopic dimensions attached to an s+p-wave 
superconductor on the left and a normal electrode on the right; see Fig. \ref{fig:Setup}.
The S electrode induces superconducting correlations into the F layer via the superconducting proximity effect. We assume  
that the pair potential  has  the form 
\begin{equation}
    \label{eq:gen_delta}
    \hat{\Delta}=\Delta_s+\Delta_t \vec d\cdot\vec{\sigma}\; , 
\end{equation}
where  $\Delta_{s}$ is the isotropic singlet component, independent of momentum direction on the Fermi surface. $\Delta_t$ and the  unit vector $\vec d$ describe the amplitude and direction of the p-wave triplet component \cite{balian1963superconductivity,sigrist1991phenomenological} respectively.  Here $\vec{\sigma}$ is the vector of Pauli matrices in spin space.

Two important examples of p-wave pairing are chiral p-wave pairing, for example $d(\phi) = e^{i\phi}\vec{a}$, and helical p-wave pairing, with $\vec{d}(\phi) = \cos{\phi}\vec{a}+\sin{\phi}\vec{b}$, where $\vec{a},\vec{b}$ are orthogonal unit vectors. Here $\phi$ is the angle with respect to a chosen axis. We choose this axis to be along the interface normal. Both chiral and helical superconductors are topological superconductors \cite{schnyder2008classification}. The former breaks time reversal symmetry and has chiral edge states \cite{kallin2016chiral,hillier2009evidence}, the latter preserves time-reversal symmetry and has so-called helical edge states \cite{chiu2016classification}.

To  describe spectral and transport properties of the junction  we use  the quasiclassical Green's function (GF) formalism extended to spin-dependent fields \cite{belzig1999quasiclassical,Bergeret-2005,heikkila2019thermal}. In this case,
the GF $\bar{G}(\boldsymbol{r},E)$ is an $8 \times 8$ matrix in Keldysh-Nambu-spin space, $\bar{G} = \begin{bmatrix}\check{G}^{R}&\check{G}^{K}\\0&\check{G}^{A}\end{bmatrix}$. In this notation, we represent matrices in Keldysh-Nambu-spin space with a bar ($\bar{\cdot}$), matrices in Nambu-spin space with a check ($\check{\cdot}$) accent, and matrices in spin space with a hat ($\hat{\cdot}$). In the dirty limit, the Green's function $\bar{G}$ is determined by a diffusion equation, known as the Usadel equation \cite{Usadel}:\\
\begin{equation}\label{eq:Usadel equation}
    D\nabla\cdot(\bar{G}\nabla \bar{G})+i[(E+\vec{h}\cdot\vec{\sigma})\tau_{3},\bar{G}]=0\; ,
\end{equation}
where $D$ is the diffusion constant, $E$ is the energy, $\vec{h}$ is the exchange field, $\tau_{3}$ is the third Pauli matrix in particle-hole space and $\vec{\sigma}$ is the vector of Pauli matrices in spin space. The Usadel equation, Eq. (\ref{eq:Usadel equation}), together with the normalization condition $\bar{G}^2=\bar{\mathbf{1}}$ and the boundary conditions determine the quasiclassical GF. 

The current $I$, and the differential conductance of the system $\sigma$, can be calculated from the quasiclassical GF using the following expressions:
\begin{align}
    I &= \frac{\sigma_{N}}{16e}\int_{-\infty}^{\infty} \mathrm{d}E\text{Tr}\left\{\tau_{3}(\bar{G}\nabla \bar{G})^{K}\right\}\; ,\label{eq:CurrentDef}\\
    \sigma &= \frac{\partial I}{\partial V}\; ,\label{eq:ConductanceDef}
\end{align}
where $\sigma_{N}$ is the normal state conductance and $e$ is the electron charge.

In order to solve the  Usadel equation, Eq. (\ref{eq:Usadel equation}), in the F region one  needs boundary conditions describing both interfaces. We assume that the S and N electrodes are not affected by the F, and keep their bulk properties, that is, they are treated as reservoirs. 
At  the F/N  interface we  use the 
well known Kupriyanov-Lukichev boundary condition \cite{kuprianov1988influence}, which is written as:
\begin{align}\label{eq:boundary NN}
   \bar{G}\nabla \bar{G}(x = L) = \frac{1}{\gamma_{BN}L}[\bar{G}(x = L),\bar{G}_{N}]\; .
\end{align}
Here $\bar{G}_{N}$ is the bulk normal
metal GF, that is, $\check{G}_{N}^{R} = \tau_{3}$ and its distribution function is the Fermi-Dirac distribution function. 
The transparency of the junction is parameterized by $\gamma_{BN}$, which is  proportional to the  interface resistance.  
In case of a perfectly transparent interface $\gamma_{BN}\rightarrow0$ and Eq. (\ref{eq:boundary NN}) is equivalent to the continuity of $\bar{G}$ at this interface, that is, $\bar G(x=L)=\bar G_{N}$.

At the S/F interface we use the Tanaka-Nazarov boundary conditions \cite{tanaka2003circuit,tanaka2004theory}. These boundary conditions are an extension of the Nazarov boundary conditions \cite{nazarov1999novel}, which itself are a generalisation of the Kupriyanov-Luckichev boundary conditions. 
Here we use a new form of the Tanaka-Nazarov boundary conditions \cite{tanaka2021phys}, which is  more suited towards s+p-wave superconductors. Defining $\phi$ as the injection angle with respect to the interface  normal vector, the boundary condition reads:
\begin{equation}\label{eq:Tanaka-Nazarov}
    \bar{G}\nabla \bar{G}(x=0) = \frac{1}{\gamma_{BS} L}\langle \bar{S}(\phi)\rangle \; ,
\end{equation}
where
\begin{align}
    \bar{S}(\phi) &= \Tilde{T}(1+T_{1}^{2}+T_{1}(\bar{C}\bar{G}+\bar{G}\bar{C}))^{-1}(\bar{C}\bar{G}-\bar{G}\bar{C})\; ,\\
    \bar{C} &=\bar{H}_{+}^{-1}(\bar{\mathbf{1}}-\bar{H}_{-})\; ,\\
    \bar{H}_{+}&=\frac{1}{2}(\bar{G}_{S}(\phi)+\bar{G}_{S}(\pi-\phi))\label{eq:Hplus}\; ,\\
    \bar{H}_{-}&=\frac{1}{2}(\bar{G}_{S}(\phi)-\bar{G}_{S}(\pi-\phi))\label{eq:Hmin}\; .
\end{align}
Here we use the notation $\langle\cdot\rangle$ to denote angular averaging over all modes that pass through the interface, $\gamma_{BS} = R_{B}/R_{d}$ is the ratio of the boundary resistance to the resistivity of the F bar in the absence of a proximity effect,  $T_{1} = \Tilde{T}/(2-\Tilde{T}+2\sqrt{1-\Tilde{T}})$, and $\Tilde{T}$ is the interface transparency given by 
\begin{align}\label{eq:zdef}
\Tilde{T}(\phi) = \frac{\cos^{2}\phi}{\cos^{2}{\phi}+z^{2}}\; ,
\end{align} 
where $z$ is the BTK parameter \cite{blonder1982transition}, characterizing the strength of the barrier. It is assumed that the Fermi surface mismatch is negligible, that is, that the magnitude of the Fermi momentum is of similar magnitude in the superconductor and ferromagnet.
If $z = 0$, there is no barrier. In that case the junction is highly transparent, and there is no reflection for any mode. On the other hand, if $z$ is large, the barrier is strong and the boundary has a low transparency. 

In Eqs. (\ref{eq:Hplus}) and (\ref{eq:Hmin}) $\bar{G}_{S}(\phi)$ is the Green's function of  a bulk BCS superconductor with pair potential given by Eq. (\ref{eq:gen_delta}). We parameterized the pair potentials as  
\begin{align}
\label{eq:delta1}
    \hat{\Delta}(\phi) = \Delta_{0}\left(\frac{1}{\sqrt{r^{2}+1}}+\frac{r}{\sqrt{r^{2}+1}}\vec{d}(\phi)\cdot\vec{\sigma}\right)\; ,
\end{align}
where $\Delta_{0}$ is the energy scale of the superconducting potential, $r = \frac{\Delta_{t}}{\Delta_{s}}$ the mixing parameter, and $\vec{d}(\phi)$ is the orientation of the angular dependent d-vector. The matrix pair potential, Eq. (\ref{eq:delta1}),  has two eigenvalues, which are both independent of $\phi$, given by
\begin{align}
\Delta_{\pm} = \Delta_{0}\frac{1 \pm r}{\sqrt{r^{2}+1}}\; .
\end{align}
In the dirty limit only triplet components with d-vector parallel to $\langle \vec{d}\rangle$ are induced by the superconductor due to angular averaging \cite{kokkeler2022phys}. This can be understood as follows, because of the high rate of scattering the contributions of all modes are mixed, and thus only the angular average remains.

 Here we focus on a  helical p-wave superconductor with  $\vec{d}(\phi) = (\cos{\phi},\sin{\phi},0)$.    We have also checked  that in  the chiral case similar results  hold. For the helical pair potential, $\langle\vec{d}\rangle$ points in the $x$-direction, that is, in the same direction as the direction of the current. 
 Since the Usadel equation is unaltered by a change of spin basis, our results are equally valid for 
any other pair potential with a d-vector of the form $\vec{d}(\phi) = \cos{\phi}\vec{a}+\sin{\phi}\vec{b}$, where $\vec{a},\vec{b}$ are orthogonal unit vectors. Since  there is no orbital effect, the results only depend on the angle between $\langle \vec{d}\rangle$ and $\vec{h}$, and not on the angle between $\vec{h}$ and the direction of current.

The solution of  the retarded part of Eq. (\ref{eq:Usadel equation}) provides information about the spectral properties. For the computation of $\sigma$ one also needs  to obtain the Keldysh component of the GF. 
From the normalization condition  the Keldysh component can be  written as  $\check{G}^{K} = \check{G}^{R}\check{f}-\check{f}\check{G}^{A}$, in which the matrix structure of $\check{f}$ is given by
\begin{equation}
    \check{f} = f_{L}+f_{T}\tau_{3}+\sum_{i = 1}^{3}(f_{Ti}
    +f_{Li}\tau_{3})\sigma_{i}\; . 
\end{equation}
and satisfies the following equation:
\begin{equation}
    D\nabla\cdot(\nabla\check{f}-G^{R}\nabla \check{G}^{A}) = \check{G}^{R}[\tau_{3}\vec{h}\cdot\vec{\sigma},\check{f}]-[\tau_{3}\vec{h}\cdot\vec{\sigma},\check{f}]\check{G}^{A}.
\end{equation}
In the  electrodes one assumes that the system is in equilibrium such that  $f_{L,T}(E) = \frac{1}{2}\Big(\tanh{\frac{E+eV}{2T}}\pm\tanh{\frac{E-eV}{2T}}\Big)$ \cite{belzig1999quasiclassical}, where $V$ is voltage and $T$ is temperature of the corresponding electrode.

In the following section we show the results obtained by solving numerically  the Usadel equation, Eq. (\ref{eq:Usadel equation}), together with the boundary conditions Eqs. (\ref{eq:boundary NN}) and (\ref{eq:Tanaka-Nazarov}) in the  SFN configuration. From the knowledge of the GF we calculate the differential conductance given by Eqs. (\ref{eq:CurrentDef}) and (\ref{eq:ConductanceDef}).

\section{Differential conductance of the SFN junction}\label{sec:results}

In this section, we study the differential conductance for different magnitudes and directions of the exchange field, and for two superconducting 
regimes: the  s-wave dominated or p-wave dominated cases, corresponding to $r<1$ and $r>1$ respectively. 

We first focus on the spectral properties of the F layer.  The superconducting correlations in F, induced by the proximity effect, have the  general matrix form: 
\begin{align}
\hat{F}&=F_{0}\hat{\mathbf{1}}+F_{h} \vec{m}\cdot\vec{\sigma}+F_{d}\vec{d}_{\perp}\cdot\vec{\sigma}\; , 
\end{align}
where $F_0$ is the singlet component whereas the 
other two are triplet components, either induced by the exchange field in F or by the proximity effect. In the equation above, $\vec{m}$ is a unit vector pointing in the direction of the exchange field,  and $\vec{d}_{\perp}$ is a unit vector in the direction of $\langle\vec{d}\rangle-(\langle\vec{d}\rangle\cdot\vec{m})\vec{m}$. If $\vec{d}$ and $\vec{m}$ are parallel this term is absent.

It is instructive to linearize the Usadel equation assuming a weak proximity effect. In this case the pair amplitudes obey the following linear differential  equations:
\begin{align}
   D\nabla^{2}(F_{0}\pm F_{h}) &= 2i(E\pm h)(F_{0}\pm F_{h}),\label{eq:F1}\\
   D\nabla^{2}F_{d}&=2iE F_{d}\; .\label{eq:F2}
\end{align}
The first equation reflects the singlet - (short range) triplet conversion via the exchange field known in ferromagnets \cite{konschelle2015theory}. According to Eq. (\ref{eq:F1}), $F_{0}\pm F_{h}$ decay over the magnetic   length $\xi_{F}=\sqrt{\frac{D}{2|E\pm h|}}$. In contrast, according to Eq. (\ref{eq:F2}) the triplet component orthogonal to the local exchange field,  $F_{d}$, decays over the thermal length $\xi_{E}=\sqrt{\frac{D}{2E}}$. In other words, if the exchange field and the d-vector are parallel, only $F_{0}$ and $F_{h}$ are non-zero, but, if $\vec{h}$ and $\langle\vec{d}\rangle$ are perpendicular, $F_{d}$ is non-zero, and there are long-range triplet correlations. Thus, for large enough exchange field or long enough junctions, specifically if $h$ is much larger than the Thouless energy $E_{\text{Th}} = D/L^{2}$, $F_d$  dominates the proximity effect and hence the subgap transport of the junction.

We now go beyond the linearized case and compute numerically the differential conductance of the SFN junction. We choose following interface parameters [see Eqs. (\ref{eq:Tanaka-Nazarov}-\ref{eq:zdef})]:  $\gamma_{BS} = 2$, $z = 0.75$, and we assume a perfect contact at the FN interface at $x = L$, that is $\gamma_{BN} = 0$ in Eq. (\ref{eq:boundary NN}). 
First we assume a long junction with $(\frac{L}{\xi})^2 = 50$, where $\xi = \frac{D}{2\Delta_{0}}$.  
The direction and amplitude of the exchange field is varied. It is convenient to use the so-called Riccati-parameterization \cite{schopohl1995quasiparticle}. The Riccati parameterization and resulting equations are discussed in appendix \ref{sec:Retardedpar}. The solution method for the distribution functions is discussed in appendix \ref{sec:Keldysheq}. 

\begin{figure*}[!t]
    \centering
  {\hspace*{-2em}\includegraphics[width =8.4cm]{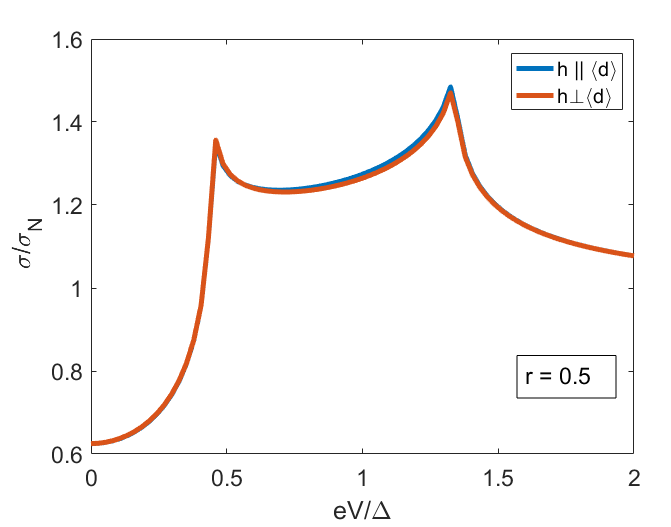}}
  \hfill
  {\hspace*{-2em}\includegraphics[width =8.4cm]{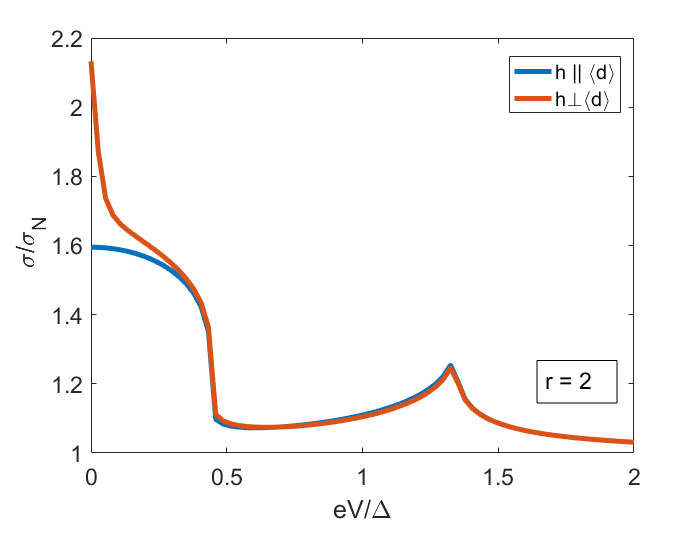}}
{\hspace*{-1.5em}\includegraphics[width =8.4cm]{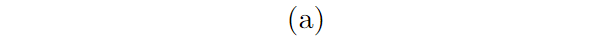}}
\hfill
{\hspace*{-2em}\includegraphics[width =8.4cm]{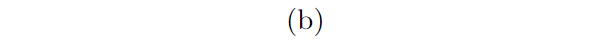}}
    \caption{The differential conductance in the SFN junction for (a) singlet dominant ($r = 0.5$) pair potential and (b) triplet dominant ($r = 2$) pair potential and  different orientations of the exchange field $h = 10\Delta_{0}$.  For both panels $L/\xi = 10$, $\gamma_{BS}=2$ and $z = 0.75$ are used. If the exchange field is parallel to the average d-vector the zero bias conductance peak (ZBCP) for triplet dominant pair potentials ($r>1$) is highly suppressed, whereas this is not the case if the exchange field is perpendicular to the d-vector.}
    \label{fig:Conductancedependence1}
\end{figure*}

The results for $\vec{h} \parallel \vec{d}$ and $\vec{h} \perp \vec{d}$ are shown in Fig. \ref{fig:Conductancedependence1} for different values of the mixing parameter $r$.
\begin{figure*}[!t]
    \centering
  {\hspace*{-2em}\includegraphics[width =8.4cm]{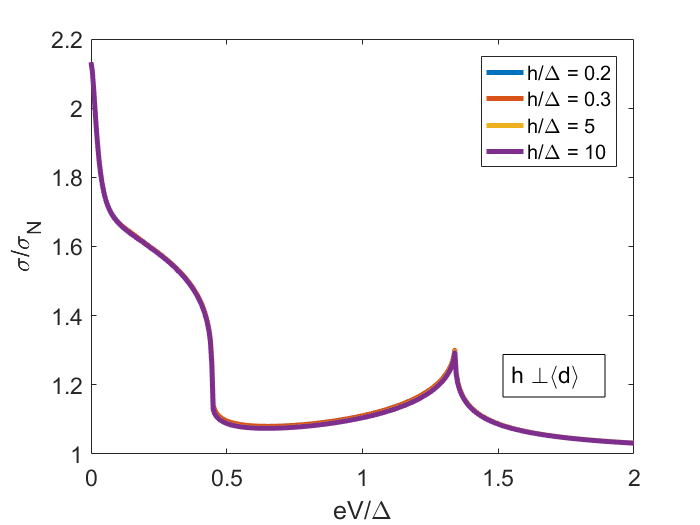}}
\hfill
  {\hspace*{-2em}\includegraphics[width =8.4cm]{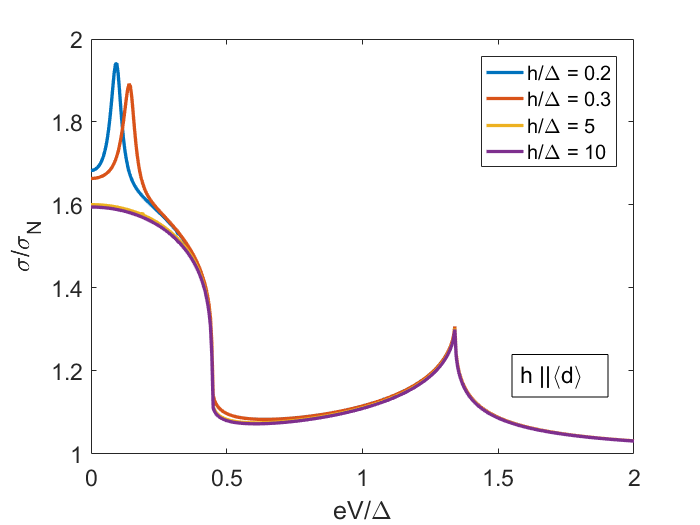}}
  {\hspace*{-1.5em}\includegraphics[width =8.4cm]{figures/A.png}}
\hfill
{\hspace*{-2em}\includegraphics[width =8.4cm]{figures/B.png}}
    \caption{$\sigma(V)$ curves for different values of  the exchange field, for the triplet dominant case $r = 2$ for (a) perpendicular and (b) parallel exchange fields.  In both panels  we choose $L/\xi = 10$, $\gamma_{BS}=2$ and $z = 0.75$. }
    \label{fig:Conductancedependence2}
\end{figure*}
There is a clear difference between the s-wave dominated and p-wave dominated pair potential junction. In the s-wave dominated case, Fig. \ref{fig:Conductancedependence1}(a), there is no ZBCP and the dependence of the differential conductance on the direction of the exchange field is weak.
In the p-wave dominated case, Fig \ref{fig:Conductancedependence1}(b), there is a ZBCP, with both a dome-like peak and a sharp peak. The dome-like peak has a width of the order of $\Delta_{0}$ and is due to surface Andreev bound states (SABS) \cite{tanaka1995theory,tanaka2018surface}. On the other hand the sharp peak has a width of the order of the Thouless energy. Such sharp peaks can also appear in systems with conventional superconductivity \cite{volkov1993proximity}. The sharp zero bias conductance peak is significantly suppressed when $\vec{h}$ is parallel to the d-vector, but not suppressed when $\vec{h}$ is perpendicular to the d-vector. 
 The anisotropy in the response to an exchange field implies that the setup can  be used to detect the presence of triplet pairing, and also  to find the direction of the d-vector of the triplet pairing.

\begin{figure*}[!t]
    \centering
   {\hspace*{-2em}\includegraphics[width =8.4cm]{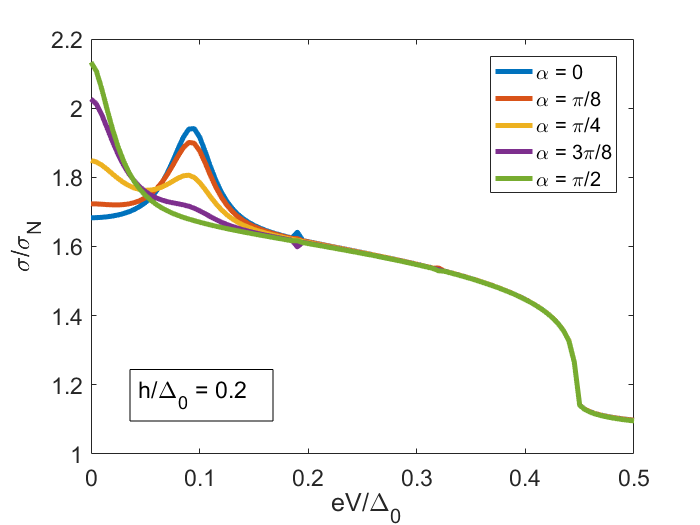}}
\hfill
  {\hspace*{-2em}\includegraphics[width =8.4cm]{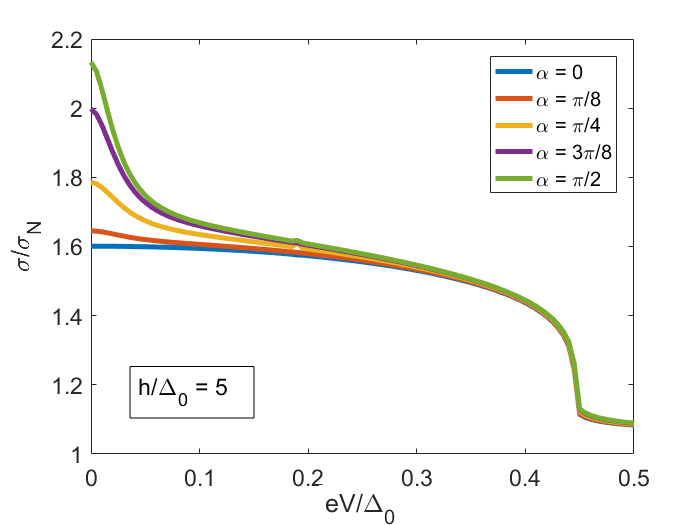}}
  {\hspace*{-1.5em}\includegraphics[width =8.4cm]{figures/A.png}}
\hfill
{\hspace*{-2em}\includegraphics[width =8.4cm]{figures/B.png}}
    \caption{$\sigma(V)$ curves for different  orientations of  the exchange field,  and $h = 0.2\Delta_{0}$ (a) and $h = 5\Delta_{0}$ (b). 
     For both panels  $r = 2$,  $L/\xi = 10$, $\gamma_{BS}=2$, and $z = 0.75$ are used.  }
    \label{fig:dIdV0}
\end{figure*}
\begin{figure}[!t]
    \centering
    \includegraphics[width =8.4cm]{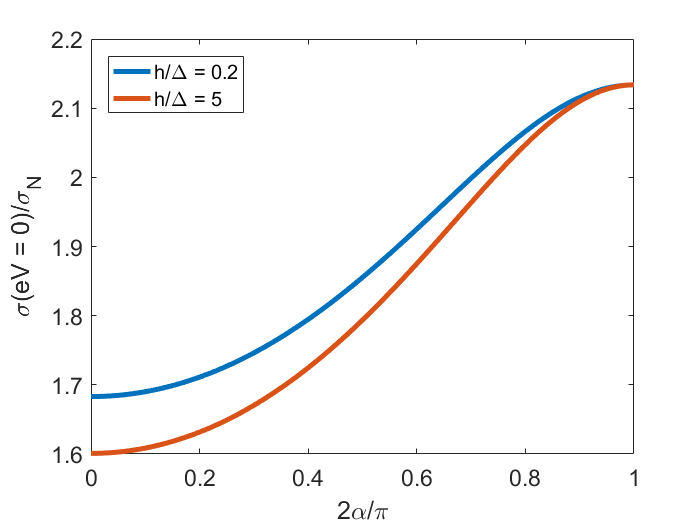}
    \caption{The magnitude of the zero bias conductance relative to the normal state conductance as a function of the direction of the exchange field for a weak ($h = 0.1\Delta_{0}$) and a strong ($h = 10\Delta_{0}$) exchange field for $r=2$. Other  parameters were set to $L/\xi = 10$, $\gamma_{BS}=2$ and $z = 0.75$.}
    \label{fig:ZBCP}
\end{figure}

To investigate the effect of the exchange field in more detail, the dependence of $\sigma$ on the strength of the exchange field for a triplet dominated pair potential ($r = 2$) is shown in Fig. \ref{fig:Conductancedependence2}. 
If the exchange field is perpendicular to the d-vector, Fig. \ref{fig:Conductancedependence2}(a), the differential conductance has only a very small dependence on the strength of the exchange field. If however, the exchange field is parallel to the d-vector, Fig. \ref{fig:Conductancedependence2}(b), for low exchange fields the sharp peak in $\sigma$ is shifted towards $eV\sim h$ and lowered, whereas the dome-like ZBCP is unaffected. As the exchange field is increased only a dome-like peak remains, without the sharp contribution. The differential conductance for $eV\in(|\Delta_{-}|,\Delta_{+})$ is also slightly affected by the strength of the exchange field, but this effect is orders of magnitude smaller than the angular dependence of the zero bias conductance. 

As the exchange field is rotated, the differential conductance varies between these two extremes in a continuous fashion. In Fig. \ref{fig:dIdV0} we show the $\sigma(eV)$ curves for different values of the angle $\alpha$ between the exchange field direction and the d-vector. We focus on the triplet dominant case, $r = 2$, for weak, Fig. \ref{fig:dIdV0}(a),  and strong, Fig. \ref{fig:dIdV0}(b), exchange field.  
For small exchange fields, Fig. \ref{fig:dIdV0}(a), a sharp peak at a nonzero voltage $eV\sim h$ develops as the angle $\alpha$ between $\vec{h}$ and $\langle\vec{d}\rangle$ is decreased. If $\alpha\approx\frac{\pi}{4}$ there is a double peak structure. As $\alpha$ decreases towards zero the ZBCP disappears. For large exchange fields, Fig. \ref{fig:dIdV0}b) no second peak appears, a decrease of $\alpha$ only leads to a suppression of the zero bias conductance.
\begin{figure*}
\centering
 {\hspace*{-2em}\includegraphics[width =8.4cm]{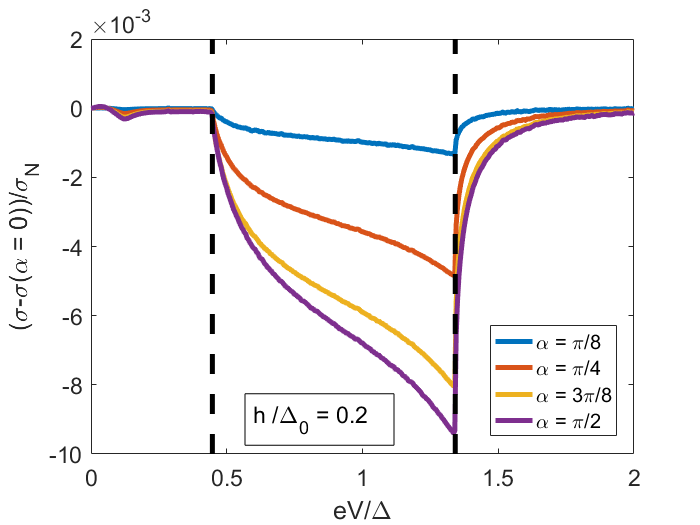}}
\hfill
  {\hspace*{-2em}\includegraphics[width =8.4cm]{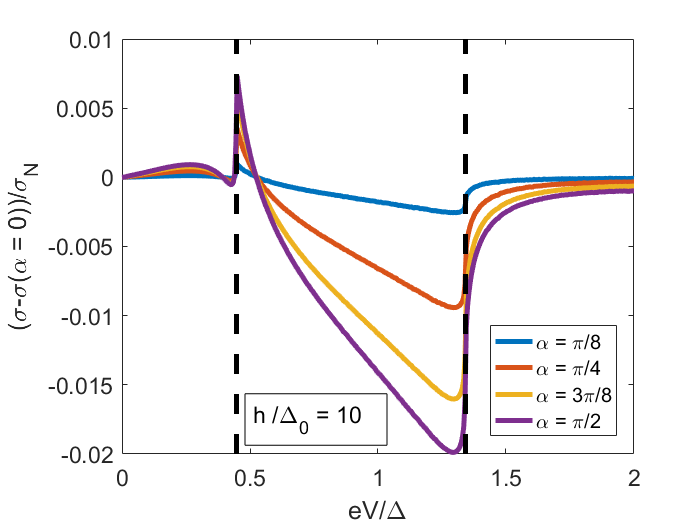}}
  {\hspace*{-1.5em}\includegraphics[width =8.4cm]{figures/A.png}}
\hfill
{\hspace*{-2em}\includegraphics[width =8.4cm]{figures/B.png}}
    \caption{The voltage dependence of $\sigma-\sigma(\alpha = 0)$ for different values of the angle $\alpha$ for a singlet dominant junction ($r = 0.5$) and for $h = 0.2\Delta_{0}$ (a) and $h = 10\Delta_{0}$ (b). The anisotropy is largest for $\Delta_{-}<eV<\Delta_{+}$, as indicated with dashed lines. In both panels $L/\xi = 10$, $\gamma_{BS}=2$ and $z = 0.75$. 
    \label{fig:R0p5angledep}}
\end{figure*}
 The angular dependence of the zero bias conductance for $r = 2$ is shown  in Fig. \ref{fig:ZBCP}. As the exchange field is rotated from a perpendicular orientation towards a parallel orientation the zero bias conductance decreases monotonically. The effect is stronger if the exchange field is increased. 

For the singlet dominated case the exchange field dependence of the differential conductance is significantly weaker, but present. To highlight the angular dependence of $\sigma$, we show the change in $\sigma$ when rotating the direction of the exchange field for $r = 0.5$ (Fig. \ref{fig:R0p5angledep}). Specifically, we show $\sigma-\sigma(\alpha =0)$ as a function of voltage for several different values of $\alpha$. Results for $h/\Delta_{0} = 0.2$ are shown in Fig. \ref{fig:R0p5angledep}(a), and results for $h/\Delta_{0} = 10$ are shown in Fig. \ref{fig:R0p5angledep}(b). Notably, we find a sizable angular dependence of the differential conductance in the range $|\Delta_{-}|<eV<\Delta_{+}$. The presence of this regime is indicative for the presence of a mixed potential, since it is absent for $r = 0$ and $r = \infty$. Moreover, the anisotropy is very small for $eV<\Delta_{-}$ and decays sharply if $eV$ is increased above $\Delta_{+}$. This means that the results can be used to infer $\Delta_{\pm}$ as illustrated by the dashed lines at $eV = |\Delta_{\pm}|$ in Fig. \ref{fig:R0p5angledep}. From this the mixing parameter $r$ can be calculated.
Thus, measuring the differential conductance  provides a way to  estimate both the direction of the d-vector and the value of the mixing parameter $r$.  

\begin{figure*}[!t]
\centering
{\hspace*{-2em}\includegraphics[width =8.4cm]{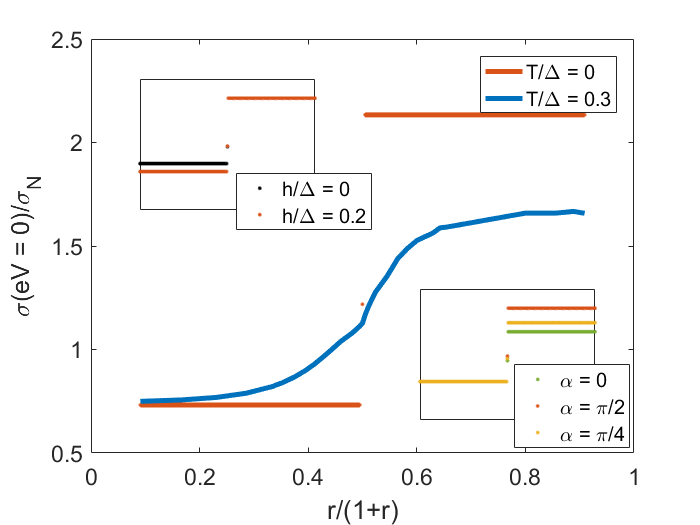}}
\caption{The zero bias conductance peak as a function of the ratio $r$ of the magnitude of the singlet and triplet components of the pair potential for zero (red curve) and finite temperature (blue curve) for a perpendicular field with $h/\Delta_{0} = 0.2$. At zero temperature there is a discontinuity, hinting towards a phase transition.
Insets: The dependence of the zero temperature zero bias conductance on the angle between the exchange field and d-vector is independent of the particular value ratio $\Delta_{t}/\Delta_{s}$. 
Other parameters are  $L/\xi = 10$, $\gamma_{BS}=2$ and $z = 0.75$.}
    \label{fig:Conductancedependencer}
\end{figure*}

The zero bias conductance in SNN s+helical p-wave junctions has another interesting feature \cite{kokkeler2022phys}. For superconductors of this type,
the zero bias conductance is independent of the particular value of $r$, it only depends on whether $r>1$ or $r<1$. We show in Fig. \ref{fig:Conductancedependencer} that this property still holds in the presence of an exchange field, that is, also the exchange field dependence is independent of the particular value of $r$.
 This can be understood as follows. The mixing parameter $r$ only enters through the Tanaka-Nazarov boundary condition, Eq. (\ref{eq:Tanaka-Nazarov}). Since the exchange field does not enter this boundary condition, its effect on the zero bias conductance is independent on the particular value of $r$.
The sharp distinction between the two regimes suggests the presence of a quantum phase transition at $r = 1$ between singlet dominated and triplet dominated superconductivity. For $T\neq0$ the dependence of $\sigma(eV = 0)$ on $r$ becomes smooth, as shown in Fig. \ref{fig:Conductancedependencer}. 

\begin{figure*}[!t]
    \centering
  {\hspace*{-2em}\includegraphics[width =8.4cm]{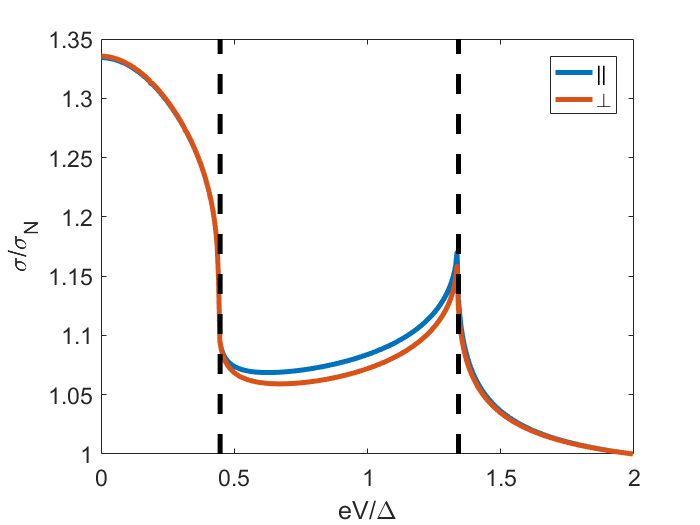}}
\hfill
  {\hspace*{-2em}\includegraphics[width =8.4cm]{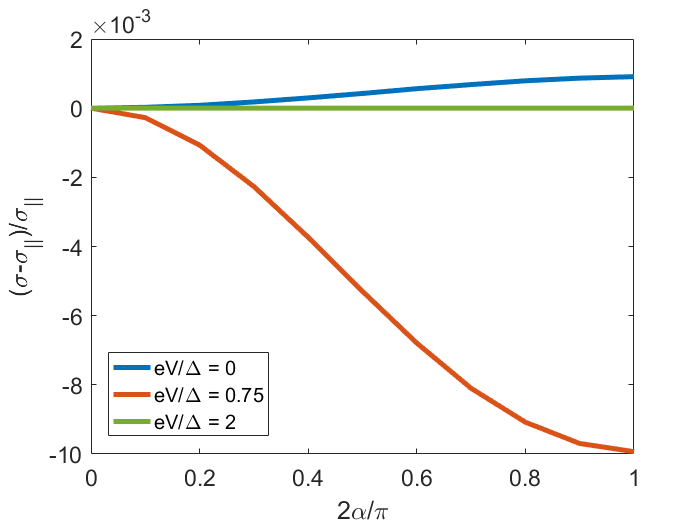}}
  {\hspace*{-1.5em}\includegraphics[width =8.4cm]{figures/A.png}}
\hfill
{\hspace*{-2em}\includegraphics[width =8.4cm]{figures/B.png}}
    \caption{Differential conductance of a short  SFN junction for the triplet dominant case,  $r = 2$ and $h = 10\Delta_{0}$. 
    (a) $\sigma(V)$ curves  for $h\perp d$ and $h\parallel d$.   
    (b) differential conductance as a function of the angle $\alpha$ for different voltages in the three regimes defined by $|\Delta_{-}|\approx 0.5$ and $\Delta_{+} \approx 1.5$. The regimes are indicated by dashed lines at $eV = |\Delta_{\pm}|$. For both panels $L/\xi = 1$, $\gamma_{BS}=2$, $z = 0.75$ and $\lambda/\xi = 10$ are used.}
    \label{fig:Short junction}
\end{figure*}

The results for junctions with an s-wave dominated pair potential (Fig. \ref{fig:R0p5angledep}) show that the anisotropy in the differential conductance for nonzero voltages can be used to determine the mixing parameter. For the long junction with p-wave dominated superconductors (Fig. \ref{fig:Conductancedependence2}) however, the anisotropy of the zero bias conductance is much larger than the anisotropy in the range $|\Delta_{-}|<eV<\Delta_{+}$ and thus the mixing parameter is hard to determine. Therefore, a shorter SF junction ($L = \xi$) with a low transparent barrier ($\gamma_{BN} = 10$) is investigated.  
In that case the Thouless energy is large and  there is no sharp peak, as shown in Fig. \ref{fig:Short junction} for $r = 2$. Only the dome-like peak remains, which has a much weaker dependence on the exchange field. The angular dependence of the conductance is largest for $|\Delta_{-}|<eV<\Delta_{+}$.
The results for $h/\Delta_{0} = 0.2$ are shown in Fig. \ref{fig:Short junction}(a).  In Fig \ref{fig:Short junction}(b) it is shown that this angular dependence is monotonic and $\sigma$ is maximized if the d-vector and exchange field are parallel. 
This is in contrast to the zero bias conductance, which is maximized if the d-vector and exchange field are perpendicular.
This difference in sign compared to the anisotropy of the zero bias conductance peak can be used as a verification. Therefore, $|\Delta_{-}|$ and $\Delta_{+}$, and thus the mixing parameter $r$ can be determined accurately, as indicated by the dashed lines in Fig. \ref{fig:Short junction}(a).

\section{Discussion and Conclusions}\label{sec:conclusion}
We have shown that for an SFN junction an electrical measurement, namely the differential conductance, can be used to identify s+p-wave pairing, and to distinguish different types of s+p-wave superconductors. The Keldysh-Usadel equation together  with  the Tanaka-Nazarov boundary conditions have been used to calculate the differential conductance,  $\sigma$, for a junction between an s+p-wave superconductor and a ferromagnetic metal.  
We have found that the  $\sigma(eV)$ curves   depend  on  both the relative strength of the singlet and triplet components and the direction of the exchange field. If the exchange field is parallel to the d-vector of the s+p-wave superconductor, the zero bias conductance peak is suppressed and a finite bias peak appears. On the other hand, if the exchange field is perpendicular to the injected spins the zero bias conductance peak is independent of the exchange field strength. 

Thus, the experiment that we propose based on our calculation  provides a tool to characterize the pair potential of superconductors, using only electrical measurements.
This implies that the dependence of the zero bias conductance peak can be easily extracted from the results.

Our results can be used to determine not only the direction of the d-vector, but also the mixing parameter $r$. Therefore, by doing the experiment modelled in our paper, the pair potential of mixed potential superconductors can be fully characterized.

We found that both the regimes $h\gg\Delta_{0}$ and $h<\Delta_{0}$ are of interest.  Ferromagnets like Fe, Co, or Ni  have exchange fields  typically much larger than the critical temperatures of superconductors, and thus, they can be used for the regime $h\gg\Delta_{0}$. To access the regime $h<\Delta_{0}$ as well, one can use a thin normal metal layer, proximized by a ferromagnetic insulator \cite{zhang2020phase,Hijano-2021}.

Our method can be generalized to study more general types of mixed-parity superconductors, including the possibility of d-wave or f-wave pair potentials. 

\section*{Acknowledgements}
We thank S. Ili\'c, A.A. Golubov and Y. Tanaka for fruitful discussions.
We  acknowledge  financial support from Spanish AEI through project PID2020-114252GB-I00 (SPIRIT), the Basque Government through grant IT-1591-22, and  European Union’s Horizon 2020 Research and Innovation Framework Programme under Grant No. 800923 (SUPERTED). A.H. acknowledges funding by the University of the Basque Country (Project PIF20/05). F.S.B. thanks
Prof. Björn Trauzettel for his hospitality at Würzburg
University, and the A. v. Humboldt Foundation for
financial support.

\vspace{1cm}
\onecolumngrid

\appendix
\numberwithin{equation}{section}
\renewcommand{\thesubsection}{\arabic{subsection}}

\section{Weak proximity effect}\label{sec:Analytic Results}
In this appendix, we present analytic results obtained in the limit of a weak proximity effect. We show that in the case of a field perpendicular to the d-vector there are long range triplet correlations \cite{Bergeret-2005}, whereas in the case of a parallel field all triplet correlations are short-range. These results indicate why the properties of the junction are different for different orientations of the exchange field with respect to the direction of the d-vector. The applied formalism can only be used for the retarded part, since the effect of the superconductor on the differential conductance is of second order in the pairing amplitudes and is thus ignored in this limit.

If the proximity effect in the junction is small, that is, the pair amplitudes are small compared to the density of states, the following approximation can be made:
\begin{equation}
    \check{G}^{R} \approx \begin{bmatrix}
    \hat{\mathbf{1}}&\hat{F}\\-\hat{\Tilde{F}}&-\hat{\mathbf{1}}
    \end{bmatrix}\; ,
\end{equation}
where $\hat{F},\hat{\Tilde{F}}$ are the pair amplitudes. This parameterization satisfies the normalisation condition up to first order. Introducing $\vec{d}_{\perp}$ as a unit vector in the direction of $\langle\vec{d}\rangle-(\vec{m}\cdot\langle\vec{d}\rangle)\vec{m}$, where the notation $\vec{m}$ is used to denote a unit vector in the direction of $\vec{h}$, $\hat{F}$ can be decomposed into
\begin{align}
    \hat{F} &= F_{0}\hat{\mathbf{1}}+F_{h}\vec{m}\cdot\vec{\sigma}+F_{d}\vec{d}_{\perp}\cdot\vec{\sigma}+F_{dh}(\vec{d}_{\perp}\times\vec{m})\cdot\vec{\sigma}\; ,\\
    F_{\pm} &= F_{0}\pm F_{h}\; ,
\end{align}
and we decompose the components of $\hat{\Tilde{F}}$ analogously. Note that this decomposition does not use any additional assumption, it is  general for matrices in $\mathbb{C}^{2\text{x}2}$. The following equations are satisfied:
\begin{align}
    D\nabla^{2}F_{\pm} = 2i(E\pm h)F_{\pm}\label{eq:linearisedUsadel}\; ,\\
    D\nabla^{2}F_{h,dh} = 2iEF_{h,dh}\; .\label{eq:linFdFdh}
\end{align}
Taking into account that $F(x = L) = 0$ due to the good contact with the normal metal reservoir at $x = L$, the solutions to Eqs. (\ref{eq:linearisedUsadel}) and (\ref{eq:linFdFdh}) read:
\begin{align}
    F_{\pm} = C_{\pm}\sinh{\sqrt{\frac{2i(E\pm h)}{D}}(L-x)}\; ,\\
    F_{d,dh} = C_{d,dh}\sinh{\sqrt{\frac{2iE}{D}}(L-x)}\; .\label{eq:FdFdh}
\end{align}
Now, at $x = 0$, the relation $\check{G}^{R}\nabla \check{G}^{R} = \frac{1}{\gamma_{BS}L}[\check{C}^{R},\check{G}^{R}]$ should be satisfied, where $\check{C}^{R}$ is the retarded part of $\bar{C}$, the boundary term presented in the main text.

The pair amplitudes $F_{\pm}$ have a decay length $\xi_{F} = \sqrt{\frac{D}{|E\pm h|}}$, whereas the the pair amplitudes $F_{d,dh}$ decay over a length $\xi_{E} = \sqrt{\frac{D}{E}}$, and are thus unaffected by the exchange field. These are the so-called long-range triplet correlations. Using the Tanaka-Nazarov boundary conditions, explicit expressions for the coefficients can be found. For clarity of notation we only show the case in which a single mode, the one at normal incidence contributes. If other modes are taken into account the notation becomes more cumbersome, but the results are very similar.

First,consider the case in which $\vec{m} = \langle\vec{d}\rangle$. In that case all spin dependent terms in the problem are proportional to $\vec{m}\cdot\vec{\sigma}$. This implies that $(\vec{m}\cdot\vec{\sigma})\check{G}(\vec{m}\cdot\vec{\sigma}) = \check{G}$. Therefore, $F_{\pm},\Tilde{F}_{\pm}$ are the only nonzero components.

The boundary conditions imply
\begin{align}
    C_{\pm} &=\frac{1}{\sinh{(\sqrt{2i\frac{E\pm h}{D}}L)}} \left(2i(E\pm h)+\frac{1}{\gamma_{BS}}\frac{2(g_{+}+g_{-})}{1+g_{+}g_{-}-f_{+}f_{-}}\right)^{-1}\frac{1}{1+g_{+}g_{-}-f_{+}f_{-}}(f_{+}+f_{-}\pm(g_{+}f_{-}-g_{-}f_{+}))\; ,\\
    \Tilde{C}_{\pm} &=\frac{1}{\sinh{(\sqrt{2i\frac{E\pm h}{D}}L)}} \left(2i(E\pm h)+\frac{1}{\gamma_{BS}}\frac{2(g_{+}+g_{-})}{1+g_{+}g_{-}-f_{+}f_{-}}\right)^{-1}\frac{1}{1+g_{+}g_{-}-f_{+}f_{-}}(f_{+}+f_{-}\mp(g_{+}f_{-}-g_{-}f_{+}))\; .
\end{align}
For $h = 0$, the contributions proportional to $f_{+}+f_{-}$ have the same sign in $C_{+}$ and $C_{-}$. Therefore, they only contribute to $F_{+}+F_{-} = F_{0}$ and are singlets induced by the singlet component of the pair potential. For $h\neq 0$, the contributions induced by the singlet pair potential are partially singlets and partially triplets. On the other hand, the terms proportional to $(g_{+}f_{-}-g_{-}f_{+})$ are induced by the triplet component of the pair potential. For $h = 0$ they only contribute to $F_{+}-F_{-} = F_{h}$ and thus they are triplets, but for $h\neq 0$ they are partially singlets and partially triplets.

On the other hand, if $\vec{h}$ and $\vec{d}$ are perpendicular, the terms induced by the triplet part of the s+p-wave pair potential drop out of Eq. (\ref{eq:linFdFdh}) for $F_{\pm},\Tilde{F}_{\pm}$, and the expressions for $C_{\pm},\Tilde{C}_{\pm}$ reduce to 
\begin{align}
C_{\pm} &=\Tilde{C}_{\pm} = \frac{1}{\sinh{(\sqrt{2i\frac{E\pm h}{D}}L)}} \left(2i(E\pm h)+\frac{1}{\gamma_{BS}}\frac{2(g_{+}+g_{-})}{1+g_{+}g_{-}-f_{+}f_{-}}\right)^{-1}\frac{1}{1+g_{+}g_{-}-f_{+}f_{-}}(f_{+}+f_{-})\; .
\end{align}
Again, these terms are singlets for $h = 0$ and become partially singlets, partially triplets for $h\neq 0$.
The component $F_{d}$ is also nonzero in this case,
\begin{align}
    C_{d} &=-\Tilde{C}_{d} = \frac{1}{2iE\sinh{(\sqrt{2i\frac{E}{D}}L)}} \frac{(g_{+}f_{-}-g_{-}f_{+})}{1+g_{+}g_{-}-f_{+}f_{-}}\; .
\end{align}
Since the equation for $F_{d}$ is not mixed with the equation for $F_{0}$, these triplet correlations can not be converted to singlet correlations.
The boundary condition still has no terms entering Eq. (\ref{eq:linFdFdh}) for $F_{dh}$, and the equation for $F_{dh}$ is uncoupled from the other equations. Therefore, $F_{dh} = 0$ for a perpendicular orientation as well.

In conclusion, in the case of a parallel field, the only nonzero components of the anomalous GF are $F_{\pm}$, which decay on a length scale $\xi_F = \sqrt{\frac{D}{2|E\pm h|}}$, whereas in the case of a perpendicular field, there are long-range correlations decaying on a scale $\xi_{E} = \sqrt{\frac{D}{2|E|}}$. Thus, in the case of a perpendicular field the correlations extend over the full junction as $E\xrightarrow{}0$. This explains the strong anisotropy of the junction with respect to the exchange field.
\section{Solution procedure}
In this appendix,  we discuss the implementation of the Usadel equation using the parameterization introduced in the main text.
\subsection{Retarded equations}\label{sec:Retardedpar}
The equation for the retarded part $\check{G}^{R}$ reads
\begin{align}
    D\nabla\cdot(\check{G}^{R}\nabla\check{G}^{R})+i[(E+\vec{h}\cdot\vec{\sigma})\tau_{3},\check{G}^{R}] = 0\; .
\end{align}
The Riccati-parameterization \cite{schopohl1995quasiparticle} is as follows:
\begin{equation}\label{eq:Riccati}
    \check{G}^{R} = \begin{bmatrix}(1+\hat{\gamma}\hat{\Tilde{\gamma}})^{-1}(1 - \hat{\gamma}\hat{\Tilde{\gamma}})&2(1+\hat{\gamma}\hat{\Tilde{\gamma}})^{-1}\hat{\gamma}\\2(1+\hat{\Tilde{\gamma}}\hat{\gamma})^{-1}\hat{\Tilde{\gamma}}&-(1+\hat{\Tilde{\gamma}}\hat{\gamma})^{-1}(1-\hat{\Tilde{\gamma}}\hat{\gamma})\end{bmatrix}\; ,
\end{equation}
Inserting the parameterization in Eq. (\ref{eq:Riccati}) into the Usadel equation, Eq. (\ref{eq:Usadel equation}), we find, using a derivation similar to the one presented in \cite{jacobsen2015critical}, that the Ricatti-matrices satisfy the following equations
\begin{align}
    \nabla^{2}\hat{\gamma}-2\nabla\hat{\gamma}\cdot\hat{\Tilde{N}}\hat{\Tilde{\gamma}}\nabla\hat{\gamma} &= \frac{2\omega}{D}\hat{\gamma}+\frac{i}{D}\{\vec{h}\cdot\vec{\sigma},\hat{\gamma}\}\; , \label{eq:usadel gamma}\\
    \nabla^{2}\hat{\Tilde{\gamma}}-2\nabla\hat{\Tilde{\gamma}}\cdot \hat{N}\hat{\gamma}\nabla\hat{\Tilde{\gamma}} &= \frac{2\omega}{D}\hat{\Tilde{\gamma}}+\frac{i}{D}\{\vec{h}\cdot\vec{\sigma},\hat{\Tilde{\gamma}}\}\; . \label{eq:usadel gammatilde}
\end{align}
The boundary condition at the SN interface is
\begin{align}
    \nabla \hat{\gamma} &= \frac{1}{\gamma_{BS}L}\frac{1}{2}(\hat{\mathbf{1}}+\hat{\gamma}\hat{\Tilde{\gamma}})(\hat{I}^{R}_{S12}-\hat{I}^{R}_{S11}\hat{\gamma})\; ,\label{eq:BCgS}\\
    \nabla \hat{\Tilde{\gamma}} &= \frac{-1}{\gamma_{BS}L}\frac{1}{2}(\hat{\mathbf{1}}+\hat{\Tilde{\gamma}}\hat{\gamma})(\hat{I}^{R}_{N21}+\hat{I}^{R}_{N22}\hat{\Tilde{\gamma}})\label{eq:BCTgS}\; ,
\end{align}
where $\hat{I}^{R}_{S11} = \text{Tr}_{\tau}\frac{1}{2}(1+\tau_{3})\check{I}^{R}_{S}$, $\hat{I}^{R}_{S12} =\text{Tr}_{\tau}\frac{1}{2}(\tau_{1}+i\tau_{2})\check{I}^{R}_{S} $, $\hat{I}^{R}_{S22} = \text{Tr}_{\tau}\frac{1}{2}(1-\tau_{3})\check{I}^{R}_{S}$, $\hat{I}^{R}_{S21} =\text{Tr}_{\tau}\frac{1}{2}(\tau_{1}-i\tau_{2})\check{I}^{R}_{S} $,
where the notation $\text{Tr}_{\tau}$ has been introduced to indicate partial trace over Nambu space, and
\begin{align}
   \check{I}^{R}_{S} = \langle\Tilde{T}(1+T_{1}^{2}+T_{1}(\check{C}^{R}\check{G}^{R}(x=0)+\check{G}^{R}(x=0)\check{C}^{R}))^{-1}(\check{C}^{R}\check{G}^{R}(x=0)-\check{G}^{R}(x=0)\check{C}^{R})\rangle\; ,
\end{align}
where $\check{G}(x = 0)$ is found by substitution of $\hat{\gamma}(x=0)$ and $\hat{\Tilde{\gamma}}(x=0)$, and $T_{1}$ and $\check{C}^{R}$ are as defined in the main text.
Similarly, the boundary conditions at the boundary with the normal metal reservoir read
\begin{align}
    \nabla \hat{\gamma} &= \frac{-1}{\gamma_{BN}L}\frac{1}{2}(\hat{\mathbf{1}}+\hat{\gamma}\hat{\Tilde{\gamma}})(\hat{I}^{R}_{N12}-\hat{I}^{R}_{N11}\hat{\gamma})\; ,\label{eq:BCgN}\\
    \nabla \hat{\Tilde{\gamma}}& = \frac{1}{\gamma_{BN}L}\frac{1}{2}(\hat{\mathbf{1}}+\hat{\Tilde{\gamma}}\hat{\gamma})(\hat{I}^{R}_{N21}+\hat{I}^{R}_{N22}\hat{\Tilde{\gamma}})\label{eq:BCTgN}\; ,
\end{align}
where $\hat{I}^{R}_{N11} = \text{Tr}_{\tau}\frac{1}{2}(1+\tau_{3})\check{I}^{R}_{N}$, $\hat{I}^{R}_{N12} =\text{Tr}_{\tau}\frac{1}{2}(\tau_{1}+i\tau_{2})\check{I}^{R}_{N}$, $\hat{I}^{R}_{N22} = \text{Tr}_{\tau}\frac{1}{2}(1-\tau_{3})\check{I}^{R}_{N}$, $\hat{I}^{R}_{N21} =\text{Tr}_{\tau}\frac{1}{2}(\tau_{1}-i\tau_{2})\check{I}^{R}_{N}$,
and
\begin{align}
    \check{I}^{R}_{N} = \frac{1}{\gamma_{B}L}[\check{G}(x=L),\check{G}_{N}^{R}]\; ,
\end{align}
where $\check{G}(x = L)$ can be calculated using $\hat{\gamma}(x = L)$ and $\hat{\Tilde{\gamma}}(x = L)$, and $\check{G}_{N}$ is the bulk Green's function of a normal metal as given in the main body of the article. Eqs. (\ref{eq:usadel gamma}) to (\ref{eq:BCTgN}) were solved numerically using the MATLAB built-in bvp5c.
\subsection{Keldysh equations}\label{sec:Keldyshpar}
\label{sec:Keldysheq}
In the case without an exchange field, a relatively compact analytic expression for the resistance can be found because the equations for the different spin components can be separated \cite{kokkeler2022phys}. If an exchange field is present this is not possible anymore, and the Keldysh equations for the distribution functions need to be solved. 
The Usadel equation for the distribution function $\check{f}$ reads
\begin{align}
    &D\nabla\cdot(\nabla \check{f}-\check{G}^{R}\nabla \check{f}\check{G}^{A})+D(\check{G}^{R}\nabla \check{G}^{R})\cdot\nabla \check{f}-D\nabla \check{f}\cdot(\check{G}^{A}\nabla \check{G}^{A})\nonumber\\&=-iE(\check{G}^{R}[\check{f},\tau_{3}]-[\check{f},\tau_{3}]\check{G}^{A})-i(\check{G}^{R}[\check{f},\tau_{3}\vec{h}\cdot\vec{\sigma}]-[\check{f},\tau_{3}\vec{h}\cdot\vec{\sigma}]\check{G}^{A})\; .\label{eq:Keldysh}
\end{align}
Since $\check{f}$ only has $\tau_{0}$ and $\tau_{3}$ components, the first term on the right cancels out. The second term however, does contribute, as the spin dependence of the distribution functions is non-trivial.
Using that the retarded and advanced Green's function must satisfy the retarded and advanced components of the Usadel equation, the equation can be written as
\begin{align}
    D\nabla\cdot(\nabla\check{f}-\check{G}^{R}\nabla\check{f}\check{G}^{A}) = i (G^{R}[\tau_{3}\vec{h}\cdot\vec{\sigma},\check{f}]-[\tau_{3}\vec{h}\cdot\vec{\sigma},\check{f}]G^{A})\; .
\end{align}

Taking the trace of Eq. \eqref{eq:Keldysh} results in

\begin{align}
    &D\nabla\cdot\Bigg(\nabla f_{L0}(4-\text{Tr}(\check{G}^{R}\check{G}^{A}))-\sum_{i = 1}^{3}\nabla f_{Ti}\text{Tr}(\check{G}^{R}\sigma_{i}\check{G}^{A})\Bigg)+D\sum_{i = 1}^{3}\nabla f_{i}\cdot\text{Tr}(\check{G}^{R}\nabla \check{G}^{R}\sigma_{i}-\sigma_{i}\check{G}^{A}\nabla \check{G}^{A})\nonumber\\
    &+D\nabla\cdot\Bigg(\nabla f_{T0}(-\text{Tr}(\check{G}^{R}\tau_{3}\check{G}^{A}))-\sum_{i = 1}^{3}\nabla f_{Li}\text{Tr}(\check{G}^{R}\tau_{3}\sigma_{i}\check{G}^{A})\Bigg)\nonumber\\&
    +(f_{L2}h_{3}-f_{L3}h_{2})\text{Tr}(\check{G}^{R}\tau_{3}\sigma_{x}-\tau_{3}\sigma_{x}\check{G}^{A})+(f_{T2}h_{3}-f_{T3}h_{2})\text{Tr}(\check{G}^{R}\sigma_{x}-\sigma_{x}\check{G}^{A})\nonumber\\&+(f_{L3}h_{1}-f_{L1}h_{3})\text{Tr}(\check{G}^{R}\tau_{3}\sigma_{y}-\tau_{3}\sigma_{y}\check{G}^{A})+(f_{T3}h_{1}-f_{T1}h_{3})\text{Tr}(\check{G}^{R}\sigma_{y}-\sigma_{y}\check{G}^{A})\nonumber\\&+(f_{L1}h_{2}-f_{L2}h_{1})\text{Tr}(\check{G}^{R}\tau_{3}\sigma_{z}-\tau_{3}\sigma_{z}\check{G}^{A})+(f_{T1}h_{2}-f_{T2}h_{1})\text{Tr}(\check{G}^{R}\sigma_{z}-\sigma_{z}\check{G}^{A})\nonumber\\
    &=0\; .\label{eq:Keldyshparametrised}
\end{align}
In a similar way, equations are obtained by taking the trace after multiplication by $\tau_{3}$, $\sigma_{j}$ and $\tau_{3}\sigma_{j}$ for $j = 1...3$.\\
The boundary conditions can be found in a similar way, by taking the corresponding traces over the equation
\begin{align}
    (\nabla \check{f}-\check{G}^{R}\nabla \check{f} \check{G}^{A})+\check{G}^{R}\nabla \check{G}^{R}\check{f}-\check{f}\check{G}^{A}\nabla \check{G}^{A} = \check{I}^{K}_{S/N}\; .
\end{align}
In this expression $\check{G}^{R,A}$ can be calculated directly from the retarded equation, using $\check{G}^{A} = -\tau_{3}(\check{G}^{R})^{\dagger}\tau_{3}$, and $\check{I}_{K,S/N}$ depends on both the retarded Green's function $\check{G}^{R}$ and the distribution function $\check{f}$, evaluated at $x = 0$ (for $\check{I}_{K,S}$) or $x = L$ (for $\check{I}_{K,N}$) and the Green's function in the electrode.
A set of eight non-constant coefficient second order linear differential equations are found. 
In the most general case, all coefficients can be nonzero, and analytical formulas are expansive and do not give many insights. Therefore, it was decided to solve the equations numerically using MATLAB bvp5c. The corresponding expressions for current can then be computed directly. By doing this as a function of the value of $f_{T0}$ attained at the normal metal reservoir, the current and differential conductance can be computed.

\twocolumngrid
\nocite{*}
\bibliography{biblio}

\end{document}